\documentclass[12pt]{iopart}
\usepackage{iopams,setstack,graphicx,amsthm,amssymb,color}
\usepackage{ulem}   %{\sout{texto aqui} risco horizontal
\begin{document}
\title[Binary System with different masses in the characteristic formulation]{Point Particle Binary System with Components of Different Masses in the Linear Regime of the Characteristic Formulation of General Relativity}
\author{C E Cede\~no M$^1$ , J C N de Araujo$^2$}
\address{Divis\~ao de Astrof\'isica, Instituto Nacional de Pesquisas Espaciais,
	Av. Dos Astronautas 12227-010 Jardim da Granja, S\~ao Jos\'e dos Campos SP, Brazil}
\ead{$^1$eduardo.montana@inpe.br}
\ead{$^2$jcarlos.dearaujo@inpe.br}
\begin{abstract}
A study of binary systems composed of two point particles with different masses in the linear regime of the characteristic formulation of general relativity with a Minkowski background is provided. The present paper generalizes a previous study. The boundary conditions at the world tubes generated by the particles's orbits are explored, where the metric variables are decomposed in spin-weighted spherical harmonics. The power lost by the emission of gravitational waves is computed using the Bondi News function. The power found is the well-known result obtained by Peters and Mathews using a different approach. This agreement validates the approach considered here. Several multipole terms contribution to the gravitational radiation field is also shown.
\end{abstract}
\vspace{2pc}
\\
\submitto{\CQG}
\maketitle
\section{Introduction}

The production of gravitational waves by binary systems, in the limit in which mass and momentum transfer do not occur, has been well described with high accurate precision with both the Post-Newtonian and the Post-Minkowskian approaches. These are the most useful tools in general relativity to provide good estimates of the features of the gravitational wave sources. Recently, great advances in these methods have improved their precision, using different multipole expansions. For a review of these issues we refer the reader to \cite{B14}.

On the other hand, the characteristic formulation of general relativity on outgoing null cones provides an interesting point of view for treating gravitationally radiating systems, since it is based on the radiation's coordinates. 
Originally formulated by Bondi {\it et. al.} in the 1960s \cite{BBM62,S62}, it has served as the basis for the development of recent and accurate characteristic numerical codes. These have evolved from the quasi-spherical regime, where the non-linear angular terms are neglected \cite{BGLW96}, to the full non-linear case \cite{BGLMW97}. 
With the inclusion of (perfect fluids) matter sources into the numerical codes \cite{BGLMW99}, {\it post hoc} extensions were made, putting the system of equations in terms of only first order angular derivatives \cite{G01}. 
The  development of the Cauchy-characteristic matching (CCM) method \cite{BGLMW05} allows the implementation of highly accurate gravitational wave propagation techniques. 
Great advances in the precision for the angular derivatives were made through the introduction of a gnomonic atlas to cover the angular manifold \cite{GBF07,RBLTS07}. 
Also, there has been impressive progress in spectral methods \cite{OR11} and in mixed methods, combining finite-difference schemes and spectral representation of the angular part of the field equations, including use of the Runge-Kutta 4 method to perform the time integration \cite{C13,RBP13,HS15}.
\\ The main purpose of such codes, after all, is to simulate the gravitational radiation patterns produced, for example, in stellar  collapses or by coalescing black hole$-$black hole (BH-BH), black hole$-$neutron star (BH-NS) or neutron star$-$neutron star (NS-NS) binary systems.  As calibration tools, some toy models have been constructed from analytical solutions of the Einstein field equations in the linear regime (see, e.g., \cite{B05,RBPS10,BPR11,K12}).   
\\ These toy models allow one to study the boundary conditions that the metric and their derivatives should satisfy in different situations in a simple, clear and straightforward way. Bishop {\it et. al.} \cite{B05,BPR11} and Kubeka \cite{K12}, provide a prescription to solve analytically the field equations for the vacuum in the linear regime. Then, they construct solutions for a variety of simple systems, such as a static (dynamical) thin shell, and a binary system composed of point particles of equal masses moving in circular orbits. However, the issue concerning the boundary conditions deserves additional studies, and most importantly, binaries of different masses in circular orbits have not been considered and therefore deserve  closer scrutiny.
\\ Therefore, the present paper generalizes a study presented in Sec. III of  \cite{BPR11}, which is related to binary systems. It is worth stressing that the approach considered here has been performed in a completely 	analytical way. The validity of this approach is supported by finding that the power lost in gravitational waves by the system agrees with the
well-known result obtained by Peters and Mathews \cite{PM63} using a different approach.  
\\In the next sections, the case of a binary system composed of two point particles of different masses moving in circular orbits in a Minkowski background is studied. In Sec. II, we present a brief review of the formalism, the notation used and the Einstein field equations in the characteristic formulation on null cones oriented to the future. In Sec. III the specific solutions for the field equations are constructed by expanding the metric variables in terms of the $_sZ_{lm}$ spin-weighted spherical harmonics. A discussion on the boundary conditions satisfied by the coefficients in these expansions for arbitrary values of the $l$ and $m$ modes is provided. In Sec. IV the gravitational radiation emitted by the binary system is computed using the Bondi News function. In particular, the contribution to the gravitational radiation by the ($l$) odd modes due to the asymmetry (different masses) of the system is considered, and it is shown how these terms vanish if the system is symmetric (equal masses). Finally, in Sec. V  other issues for further study are presented.

\section{Formalism}
The characteristic formulation based on future-oriented null cones, when the eth
formalism is employed, is well known and has been used to explore a wide variety of
problems \cite{GBF07,RBP13,HS15,C13,RBPS10,TBRSCKS13}. However, a brief review of some of its most important aspects is necessary to present the notation and conventions used here. We use the geometrised unit system, i.e. $G=c=1$. The Greek indices run from 1 to 4, whereas the capital latin letters run from 3 to 4, labelling the angular indices. The coordinates used are $x^{\mu}=(u,r,x^{A})$, where the retarded time $u$ labels each null cone upon which the spacetime is foliated, $r$ is the luminosity distance and $x^A$ represents the angular coordinates. Here $x^A$ can denote $(\theta,\phi)$ or $(q,p)$ for the usual spherical coordinates  or for the stereographic two patch representation, respectively. In these coordinates the Bondi-Sachs metric takes the form
\begin{eqnarray}
\label{bs}
ds^2&=&-\left(e^{2\beta}\left(1+\frac{w}{r}\right)-r^2h_{AB}U^AU^B\right)du^2 -2e^{2\beta}dudr 
\nonumber \\&& 
-2r^2h_{AB}U^Bdx^A du +r^2h_{AB}dx^Adx^B\;,
\end{eqnarray}
where $w$ and the redshift $\beta$ are related to the square of the ADM lapse function 
\cite{BGLW96}. The shift vector between two successive null cones is labelled by $U^A$. The metric associated with the angular manifold (unit sphere) is represented by $h_{AB}$ ($q_{AB}$). In addition, the Bondi gauge is imposed, i.e.  $\det(h_{AB})=\det(q_{AB})=1$. Further it is required that $h_{AB}h^{BC}=q_{AB}q^{BC}=\delta_B^{~C}$. 
\\The metric of the unit sphere is decomposed as $q_{AB}=q_{(A}\overline{q}_{B)}$, where the overline indicates complex conjugation, the parenthesis symmetrization and the dyad $q^{A}$ are the null tangent vectors along the coordinate lines associated with the angular charts used to make the finite covering of the sphere. These vectors satisfy the relations: $q_Aq^A=0$ and $q_A\overline{q}^A=2$. In stereographic coordinates the $q^A$ vectors take the form
\begin{equation*}
q^A=\frac{(1+\zeta\overline{\zeta})(\delta^A_{~~3}+i\delta^A_{~~4})}{2}\;,\hspace{0.5cm}  \zeta=\tan\left(\frac{\theta}{2}\right)e^{i\phi}\;,\hspace{0.5cm}\zeta=q+ip
\end{equation*}
(see e.g. \cite{BGLW96,BGLMW97,GPLPW97,G01}) %,BGHLW03,B05,BGLMW05,GBF07,RBLTS07,BPR11,C13}) 
 whereas in spherical coordinates (see e.g. \cite{RBLTS07,BPR11,HS15}),
\begin{equation*}
q^A=\delta^A_{~~3}+i\csc\theta \delta^A_{~~4}\;.
\end{equation*}
All tensor quantities, such as the metric or the Ricci tensors, are contracting with the dyad producing  spin-weighted scalars. The metric $h_{AB}$ for the angular manifold can be represented by $J=q^Aq^Bh_{AB}/2$ and $K=q^A\overline{q}^Bh_{AB}/2$, and $U^A$ by $U=q^AU_A$. Also, the Bondi gauge constrains $K$ to satisfy $K^2=1+J\overline{J}$. Thus $h_{AB}$ depends only on $J$. It is worth stressing that the case $J=0$ returns the symmetrical unit sphere metric.
\\Associated with the projections of the covariant differentiation referred to $q_{AB}$ onto the $q^A$ vectors, the differential operators, labelled $\eth$ and $\overline\eth$, are constructed. These operators can raise or lower the spin-weight of any spin-weighted scalar function. The spin-weighted functions are given by 
\begin{equation*}
_s\Psi=\prod_{i=1}^m \Lambda_{B_i}\prod_{j=1}^n \Lambda^{A_j}  \Psi^{B_1\cdots B_m}_{\hspace{1.1cm}A_1\cdots A_n}\;,
\end{equation*}
whereas $\eth$ is defined 
\begin{equation*}
\eth\ _s\Psi=q^D\prod_{i=1}^m \Lambda_{B_i}\prod_{j=1}^n \Lambda^{A_j}  \Psi^{B_1\cdots B_m}_{\hspace{1.1cm}A_1\cdots A_n|D}\;,
\end{equation*}
and $\overline{\eth}$ is defined
\begin{equation*}
\overline \eth\ _s\Psi=\overline{q}^D\prod_{i=1}^m \Lambda_{B_i}\prod_{j=1}^n \Lambda^{A_j}  \Psi^{B_1\cdots B_m}_{\hspace{1.1cm}A_1\cdots A_n|D}\;,
\end{equation*} 
where the vertical line in the indices indicates covariant differentiation associated with $q_{AB}$,  and the symbols $\Lambda_{B_i}$ and $\Lambda^{A_i}$ can take the values $q_{B_i}$ or $\overline{q}_{B_i}$  and $q^{A_i}$ or $\overline{q}^{A_i}$, respectively. Specifically, these operators take the following form
\begin{equation*}
\eth\ _s\Psi=q^D\ _s\Psi_{,D}+s\Omega \ _s\Psi\;, \hspace{1cm}\overline{\eth}\ _s\Psi=\overline{q}^D\ _s\Psi_{,D}-s\overline{\Omega} \ _s\Psi\;,
\end{equation*}
where $\Omega=-q^Aq^Bq_{B|A}/2$. For a complete review of these operators and their properties, see \cite{GPLPW97,NP66,GMNRS66}. 
\\The Einstein equations can be written as 
$$E_{\mu\nu}=R_{\mu\nu}-8\pi\left(T_{\mu\nu}- g_{\mu\nu}T/2\right)=0\;,$$ 
and in this formalism, they are decomposed as
\numparts
\begin{eqnarray}
&& E_{22}=0\;, \hspace{0.2cm}E_{2A}q^A=0\;, \hspace{0.2cm} \hspace{0.2cm} E_{AB}h^{AB}=0\;,\\ 
&& E_{AB}q^{A}q^{B}=0\;,\\
&& E_{11}=0\;, \hspace{0.2cm}E_{12}=0\;,\hspace{0.2cm}\hspace{0.2cm} E_{1A}q^A=0\;,
\end{eqnarray}
\endnumparts	
which corresponds to hypersurface, evolution and constraint equations, respectively \cite{BGLMW97,B05,RBPS10,BPR11,RBP13}. 
\\In the linear regime, the following set of equations corresponding to a perturbation in the Minkowski background, as originally obtained by Winicour \cite{W83} and later by Bishop \cite{B05}, is given by
\numparts
\begin{eqnarray}
&& 8 \pi  T_{22}=\frac{4 \beta _{,r}}{r}\;, \label{field_eq_1}\\
&& 8 \pi  T_{2A} q^A=\frac{\overline{\eth }J_{,r}}{2} -\eth \beta _{,r} +\frac{2 \eth \beta}{r} +\frac{\left(r^4 U_{,r}\right)_{,r}}{2r^2}\;, \label{field_eq_2}\\
&& 8 \pi  \left(h^{AB} T_{AB}-r^2 T\right) =-2 \eth \overline{\eth }\beta +\frac{\eth^2\overline{J} + \overline{\eth }^2 J}{2} +\frac{\left(r^4\left(\overline{\eth}U+\eth \overline{U}\right)\right)_{,r}}{2r^2}\nonumber\\
&& \hspace{4.0cm}+4 \beta -2 w_{,r}\;, \label{field_eq_3}\\
&& 8 \pi T_{AB}  q^A q^B=-2 \eth^2 \beta + \left(r^2\eth U \right)_{,r} - \left(r^2 J_{,r}\right)_{,r} +2 r\left(rJ\right)_{,ur}\;,\label{field_eq_4}\\
&& 8 \pi  \left(\frac{T}{2}+T_{11}\right)=\frac{\eth \overline{\eth }w}{2 r^3} + \frac{\eth \overline {\eth }\beta }{r^2} -\frac{\left(\eth \overline{U} + \overline{\eth} U \right)_{,u}}{2} +\frac{w_{,u}}{r^2} +\frac{w_{,rr}}{2 r} -\frac{2 \beta_{,u}}{r}\;, \label{field_eq_5}\\ 
&& 8 \pi  \left(\frac{T}{2}+T_{12}\right)=
\frac{\eth \overline{\eth }\beta }{r^2} -\frac{\left(r^2\left(\eth \overline{U} + \overline{\eth   }U\right)\right)_{,r}}{4r^2}
+\frac{w_{,rr}}{2 r}\;, \label{field_eq_6}\\
&& 8 \pi  T_{1A} q^A=\frac{\overline{\eth }J_{,u}}{2} -\frac{\eth^2
\overline{U} }{4} +\frac{\eth \overline{\eth }U}{4} +\frac{1}{2}\left(\frac{\eth w}{r}\right)_{,r} -\eth \beta _{,u} +\frac{\left(r^4U\right)_{,r}}{2r^2}\nonumber\\
&& \hspace{2cm}-\frac{r^2 U_{,ur}}{2} +U\;. \label{field_eq_7}
\end{eqnarray}
\label{field_eqs}
\endnumparts
\section{Binary system}
\begin{figure}[h!]
\begin{center}
\includegraphics[height=6cm]{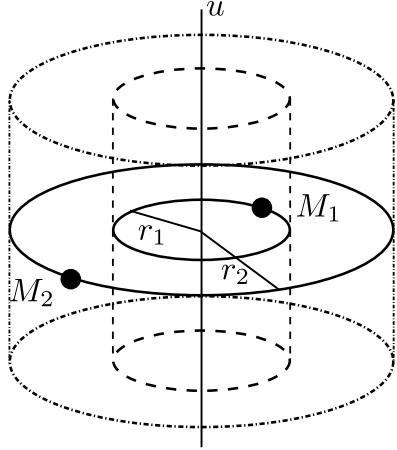}
\end{center}
\caption{Binary system with the world tubes of each orbit extended along the direction of the retarded time, separating the spacetime into three regions.}
\label{fig1}
\end{figure}

It is worth stressing that one of our aims is to study the well-known problem of a circular binary system of point particles with different masses. We show that the Peters and Mathews result for the power radiated in gravitational waves (see \ref{appendix}) can be obtained by using the characteristic formulation and the News function. Also, the present paper generalizes previous results \cite{BPR11} applying to particles with equal masses.\\
The particles held together by their mutual gravitational interaction are far enough from each other such that to first order, the interaction between them can be considered essentially Newtonian. This assumption is valid if one considers the weak field approximation, in which the Bondi-Sachs metric in stereographic null coordinates reduces to,
\begin{eqnarray}
\label{bs_lin}
\fl
ds^2&=&-\left(1-\frac{w}{r}-2\beta\right)du^2 -2(1+2\beta)dudr -2r^2\frac{(U+\overline{U})}{1+|\zeta|^2}dq du-2r^2\frac{i(U-\overline{U})}{1+|\zeta|^2}dp du
\nonumber \\
\fl && 
+2r^2\frac{\left(2+J+\overline{J}\right)}{\left(1+|\zeta|^2\right)^2}dq^2-4ir^2\frac{(J-\overline{J})}{(1+|\zeta|^2)^2}dqdp-2r^2\frac{\left(-2+J+\overline{J}\right)}{\left(1+|\zeta|^2\right)^2}dp^2\;.
\end{eqnarray} 
Note that $g_{11}\simeq -1+2\Phi$, where $\Phi=\beta+w/(2r)$ represents the Newtonian potential, as usual in this kind of approximation.  
\\We take these two particles in a Minkowski background, analogously to Peters and Mathews \cite{PM63} and Bishop {\it et. al.} in \cite{BPR11}. Such a system allows one to explore in full detail the boundary conditions across the hypersurfaces generated by their orbits shown in figure \ref{fig1}. 
\\The density that describes the binary system is given by
\begin{equation}
\rho=\frac{\delta(\theta-\pi/2)}{r^2}\bigg(M_1\delta(r-r_1)\delta(\phi-\nu u) + M_2\delta(r-r_2)\delta(\phi-\nu u-\pi)\bigg)\;,
\label{dens}
\end{equation}%
\noindent where, $r_i\ (M_i)$ are the orbital radius (mass) of each particle and $r_1 < r_2$. 
\\The orbits of the masses generate world tubes, which are extended along the retarded time coordinate, allowing the separation of the spacetime into three empty and disjoint regions: inside, between and outside the matter distribution. 
\\In order to solve equations \eref{field_eq_1}-\eref{field_eq_7} for the vacuum, the following expansion for the metric variables is used
\begin{eqnarray}
_sf=\sum_{l,m}\Re(f_{lm} e^{i|m|\nu u})\eth^s\ _0Z_{lm}\;,
\label{mfunctions}
\end{eqnarray}
where $\sum_{l,m}$ means $\sum_{l=2}^\infty\sum_{m=-l}^l$, $_sZ_{lm}$ are the spin-weighted spherical harmonics, which are defined in \cite{ZGHLW03} as $_sR_{lm}$ and are eigenfunctions of the $\left[\eth,\overline{\eth}\right]$ operator and $_sf$ represents the functions $\beta,W,U,\overline{U},J,\overline{J}$. The substitution of equation \eref{mfunctions} into \eref{field_eqs} provides a system of ordinary differential equations for the coefficients in the above expansions. The families of solutions, for $l=2$, satisfying this system of equations for the vacuum read
\numparts
\begin{eqnarray}
\beta_{2m}(r)&=&D_{1\beta 2m}\;,
\label{gen_sol_1_l2}\\
J_{2m}(r)&=&\frac{2 i D_{1\beta 2m}}{\nu  r \left| m\right| }-\frac{D_{1J2m} (\nu  r \left| m\right| -1) (\nu  r \left| m\right| +1)}{6 r^3} \nonumber \\
&& -\frac{i D_{2J2m} e^{2 i \nu  r \left| m\right| } (\nu  r \left| m\right| +i)^2}{8 \nu ^5 r^3 \left| m\right| ^5} +\frac{D_{3J2m} (\nu  r \left| m\right| -3 i)}{\nu  r \left| m\right| }\;, \label{gen_sol_2_l2}\\
U_{2m}(r)&=&\frac{2 D_{1\beta 2m} (\nu  r \left| m\right| +2 i)}{\nu  r^2 \left| m\right|
} -\frac{D_{1J2m} \left(2 \nu ^2 r^2 \left| m\right|^2 + 4 i \nu  r \left| m\right| +3\right)}{6
r^4} \nonumber\\
&& -\frac{D_{2J2m} e^{2 i \nu  r \left| m\right| } (2 \nu  r \left| m\right| +3 i)}{8 \nu ^5 r^4 \left| m\right|^5} -\frac{i D_{3J2m} \left(\nu ^2 r^2 \left| m\right| ^2+6\right)}{\nu  r^2 \left| m\right| }\;, 
\label{gen_sol_3_l2} \\
w_{2m}(r)&=& -10 r D_{1\beta 2m} +6 r D_{3J2m} (2+i \nu  r \left| m\right| ) -\frac{3 i D_{2J2m} e^{2 i \nu  r \left| m\right| }}{4 \nu ^5 r^2 \left|m\right| ^5} \nonumber\\
&&  -\frac{i D_{1J2m} ((1+i) \nu  r \left| m\right| -i)	(1+(1+i) \nu  r \left| m\right| )}{r^2}\;, \label{gen_sol_4_l2}
\end{eqnarray}
\label{gen_sol_l2}%
\endnumparts
where the constants of integration are represented by $D_{nFlm}$; here $n$ indexes the constant and $F$ corresponds to the Bondi metric function whose integration generated it. This set of families of solutions depends only on four constants, namely, $D_{1\beta 2m}$, $D_{3J2m}$, $D_{1J2m}$ and $D_{2J2m}$. This is so because the families of solutions for the coefficients $\beta_{2m}$, $J_{2m}$, $U_{2m}$ and $w_{2m}$, resulting from \eref{field_eq_1}-\eref{field_eq_4}, are constrained by \eref{field_eq_5}-\eref{field_eq_7}.  This fact is independent of $l$, thus the set of families of solutions for any $l$ will have four degrees of freedom. 
A unique solution for the whole spacetime cannot be determined by only imposing regularity of the metric variables at the null cone vertices and at the null infinity.
Therefore, additional boundary conditions must be imposed. In particular, this can be done by imposing boundary conditions on other hypersurfaces, such as in the case of the thin shells studied by Bishop \cite{B05}, in which the additional conditions are imposed across the world tubes generate by the shell itself. Once the above constants are determined, one readily obtains the metric functions $\beta$, $J$, $U$, and $w$ for the whole spacetime.
\\As divergent solutions are not expected at the vertices of the null cones, regularity conditions at these points must be imposed for the metric. In order to do so, an expansion of the metric variables around $r=0$ in power series of $r$ is made  and the divergent terms are disregarded. This procedure establishes relationships between the coefficients, leading to a family of solutions for the interior that depends only on one undetermined parameter, where in particular $\beta_{lm-}(r)=0$. One obtains, for example, for $l=2$,
\numparts
\begin{eqnarray}
\beta_{2m-}(r)= && 0\;, \label{par_sol_int_1_l2}\\
J_{2m-}(r)= && \frac{D_{2J2m-}}{24 \nu^5 r^3 \left| m\right|^5} \left(2 \nu ^3 r^3 \left| m\right|^3 - 3 i \nu^2 r^2 \left| m\right|^2 e^{2 i \nu  r \left| m\right| } -3 i \nu ^2 r^2 \left| m\right|^2   \right.\nonumber\\
&&\left. + 6 \nu  r \left| m\right|  e^{2 i \nu  r \left| m\right| } +3 i e^{2 i \nu  r \left| m\right| }-3 i\right)\;,\label{par_sol_int_2_l2}\\
U_{2m-}(r)= && -\frac{i D_{2J2m-}}{24 \nu ^5 r^4 \left| m\right| ^5}\left(2 \nu ^4 r^4 \left| m\right|^4+6 \nu ^2 r^2 \left| m\right| ^2 - 6 i \nu  r \left| m\right|  e^{2 i \nu  r \left| m\right| }  \right. \nonumber \\ 
&& \left.  -12 i \nu  r \left| m\right| +9 e^{2 i \nu  r \left| m\right| }-9\right)\;, \label{par_sol_int_3_l2}\\
w_{2m-}(r)= && \frac{D_{2J2m-}}{4 \nu^5 r^2 \left| m\right|^5} \left(2 i \nu ^4 r^4 \left| m\right|^4 + 4 \nu ^3 r^3 \left| m\right| ^3 -6 i \nu ^2 r^2 \left| m\right| ^2 -6 \nu  r \left| m\right| \right. \nonumber \\
&&\left. -3 i e^{2 i \nu  r \left| m\right| }+3 i\right)\;.  \label{par_sol_int_4_l2}
\end{eqnarray}
\endnumparts
For the intermediate region, the same structure of the general solutions is maintained, for the case $l=2$ given by \eref{gen_sol_1_l2}-\eref{gen_sol_4_l2}. That is so because there is no reason to discard any particular term, or to establish any relationship between the constants as occurs for the interior region. For the exterior region, since that is required regularity at the null infinity, the coefficient of the exponential factor $\left(\exp(2i\nu r |m|)\right)$ must be null. This means that all constants of the form $D_{2Jlm+}$, with $l=2,3,\cdots$, must be zero. Therefore, the number of degrees of freedom for the exterior family of solutions is reduced in one parameter. Thus, a family of solutions to describe the whole spacetime for the field equations \eref{field_eq_1}-\eref{field_eq_7}, with eight parameters to be determined  is obtained. Now, in order to fix these eight constants, it is necessary to impose additional boundary conditions, in particular across the time-like world tubes generated by the orbits of the particles.
\\These boundary conditions across the world tubes, i.e. when $r=r_i$, $i=1,2$ come from imposing discontinuities on the metric coefficients, 
\begin{eqnarray}
&&\left[g_{11}\right]_{r_i}=0\;, \hspace{0.2cm}\left[g_{12}\right]_{r_i}=\left.\Delta g_{12}\right|_{r_i}\;,\hspace{0.2cm}\left[g_{1A}\right]_{r_i}=0\;, \hspace{0.2cm} \left[g_{22}\right]_{r_i}=0\;,\nonumber\\
&& \left[g_{2A}\right]_{r_i}=0\;,\hspace{0.2cm}\left[g_{3\mu}\right]_{r_i}=0\;,\hspace{0.2cm}\left[g_{4\mu}\right]_{r_i}=0\;,
\label{bound_cond_1}
\end{eqnarray}
and on their first derivatives, 
\begin{eqnarray}
&&\left[g_{\mu\nu}'\right]_{r_i}=\Delta g_{\mu\nu}'\;, \hspace{0.2cm}\mu,\nu=1,\cdots 4\;,
\label{bound_cond_2}
\end{eqnarray}
where the brackets mean $[f(r)]_{r_i}=\left.f(r)\right|_{r_i+\epsilon}-\left.f(r)\right|_{r_i-\epsilon}$ . 
\\From the linearised Bondi-Sachs metric \eref{bs_lin}, and from the two sets of jump conditions \eref{bound_cond_1} and \eref{bound_cond_2}, the coefficients $\beta_{lm}, \ J_{lm},\ U_{lm}$ and $w_{lm}$ are restricted to satisfy
\begin{eqnarray}
& \left[w_{lm}(r_j)\right]=\Delta w_{jlm}\;,& \hspace{0.2cm} \left[\beta_{lm}(r_j)\right]=\Delta \beta_{jlm}\;, \nonumber\\
& \left[J_{lm}(r_j)\right]=0\;, & \hspace{0.2cm} \left[U_{lm}(r_j)\right]=0\;,
\label{bound_cond_3}
\end{eqnarray}
and for their first derivatives
\begin{eqnarray}
& \left[w'_{lm}(r_j)\right]=\Delta w'_{jlm}\;,& \hspace{0.2cm} \left[\beta_{lm}'(r_j)\right]=\Delta \beta_{jlm}'\;, \nonumber\\
& \left[J'_{lm}(r_j)\right]=\Delta J'_{jlm}\;, & \hspace{0.2cm} \left[U'_{lm}(r_j)\right]=\Delta U'_{jlm}\;,
\label{bound_cond_4}
\end{eqnarray}
where $j=1,2$, and $\Delta w_{jlm}$, $\Delta \beta_{jlm}$, $\Delta w_{jlm}'$, $\Delta \beta_{jlm}'$, $\Delta J'_{jlm}$ and $\Delta U'_{jlm}$ are functions to be determined. 
\\Solving equations \eref{bound_cond_3} and \eref{bound_cond_4}, simultaneously for both world tubes, the boundary conditions are explicitly obtained. We find that 
\numparts
\begin{equation}
\Delta \beta_{jlm}=b_{jlm}\;,\hspace{0.5cm}\Delta w_{jlm}=-2r_jb_{jlm}\;,
\label{bound_cond_5}
\end{equation}
where $b_{jlm}$ are constants. Note that this last fact implies that $\Delta \beta'_{jlm}=0$. 
We determine that the jumps for the first derivative of the $J_{lm}$ and $U_{lm}$ functions are given by
\begin{eqnarray}
&&\Delta J'_{jlm}=\frac{8\nu ^2 r_j  b_{jlm} \left| m\right|^2}{(l-1)l(l+1)(l+2)}\;,\label{bound_cond_6}\\
&&\Delta U'_{jlm}=2b_{ilm}\left(\frac{1}{r_i^2}-\frac{4i\nu |m|}{l(l+1)r_i}\right)\;.\label{bound_cond_7}
\end{eqnarray}
\endnumparts
The boundary conditions \eref{bound_cond_6} and \eref{bound_cond_7} fix all parameters of the families of solutions, providing the specific solutions for the coefficients $\beta_{lm}, \ J_{lm},\ U_{lm}$ and $w_{lm}$. These coefficients can be written as 
\begin{eqnarray}
f_{lm}(r)=& &f_{1lm}(r)\left(1-\Theta(r-r_1)\right)+f_{2lm}(r)\left(\Theta(r-r_1)-\Theta(r-r_2)\right)\nonumber\\
&&+f_{3lm}(r)\Theta(r-r_2)\;,
\label{Sols_Coeff}
\end{eqnarray}
where $f_{lm}$ represents $\beta_{lm}$, $J_{lm}$, $U_{lm}$ and $w_{lm}$, with the first subscript on the right hand side terms indicating the interior (1), the middle (2) and the exterior (3) solutions; and $\Theta(r)$ is the Heaviside function, namely
\begin{equation}
\Theta(r)=\cases{
0 & $r\le 0$\\
1 & $r>0$
}.
\end{equation}
These solutions depend explicitly on two specific parameters, namely $b_{jlm}$, with $j=1,2$, which are related to the density of matter. The specific form of these relationships is obtained by just integrating the first field equation \eref{field_eq_1} across each world tube. As a result one obtains 
\begin{equation}
b_{jlm}=2\pi r_j \rho_{jlm} \left(1+v_j^2\right)\;,
\label{betaf}
\end{equation} 
where $\rho_{jlm}$ are given by
\begin{equation}
\rho_{jlm}=\frac{1}{\pi}\int_S d(\nu u) \int_\Omega d\Omega\int_{I_j} dr \ _0\overline{Z}_{lm} e^{-i|m|\nu u}\rho\;,
\label{denscomps}
\end{equation}
in which $S = \ [0,2\pi)$, $v_j$ is the physical velocity of the particle $j$ in the space, and $I_j$ is an interval $\epsilon$ around $r_j$ that is given by $I_j=(r_j-\epsilon/2,r_j+\epsilon/2)$, with $\epsilon>0$.\\
Before proceeding, it is worth noticing that the above procedure is a generalization of Section 3 of
the paper by Bishop {\it et. al.} \cite{BPR11}, in which the binary components have equal masses. In particular, the boundary conditions are also generalized since in the present case there exist two independent world tubes. Interestingly our solution is fully analytical.
\\In order to include the null infinity, which is reached when $r$ tends to infinity, a radial compactified coordinate $s$ is defined as follows 
\begin{equation*}
s=\frac{r}{r+R_0}\;,
\end{equation*}
where $R_0$ is a compactification parameter. Thus, $0\le s\le 1$, where $s=0$ and $s=1$ corresponds to the null cone vertices and the null infinity, respectively.
\\Figure \ref{metric_functions} shows some of the coefficients of the expansion of the metric variables in terms of the compactified coordinate $s$ for $l=m=2$.
\begin{figure}[h!]
\begin{tabular}{c@{\hspace{0.7cm}}c}
\includegraphics[height=4.8cm]{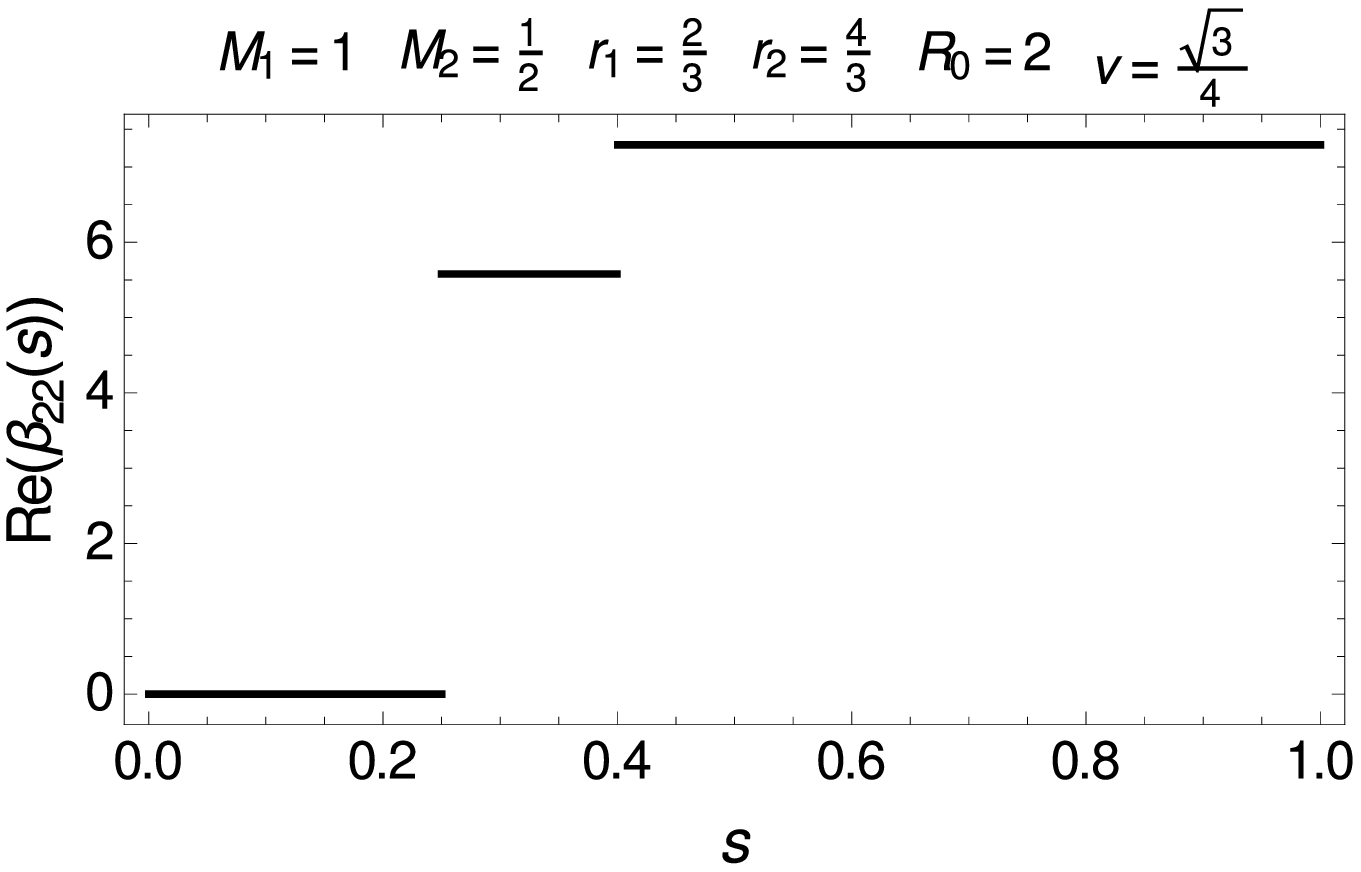} & \includegraphics[height=4.8cm]{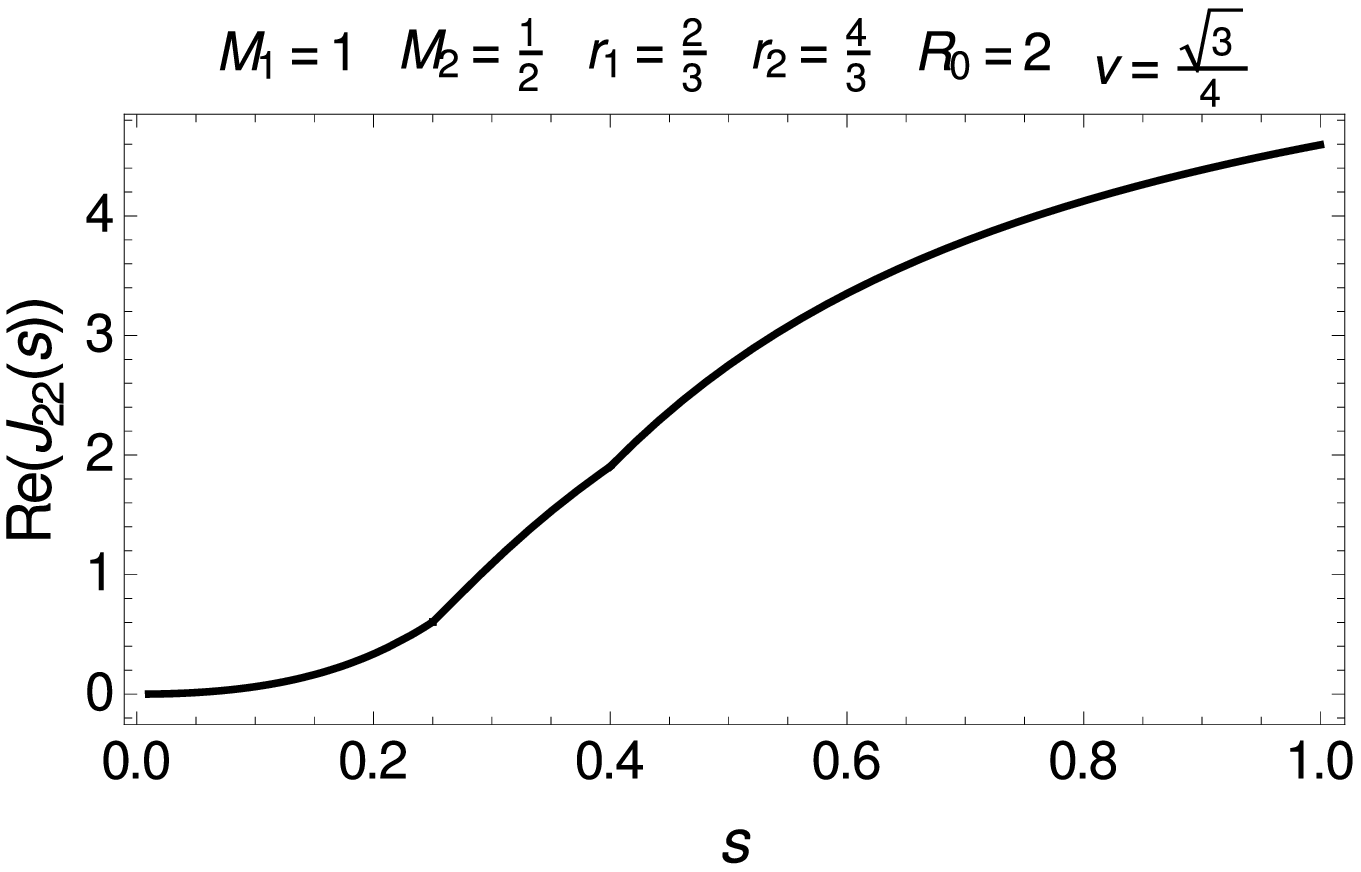}\\
(a) & (b)
\end{tabular}
\end{figure}
\begin{figure}[h!]
\begin{tabular}{c@{\hspace{0.7cm}}c}
\includegraphics[height=4.8cm]{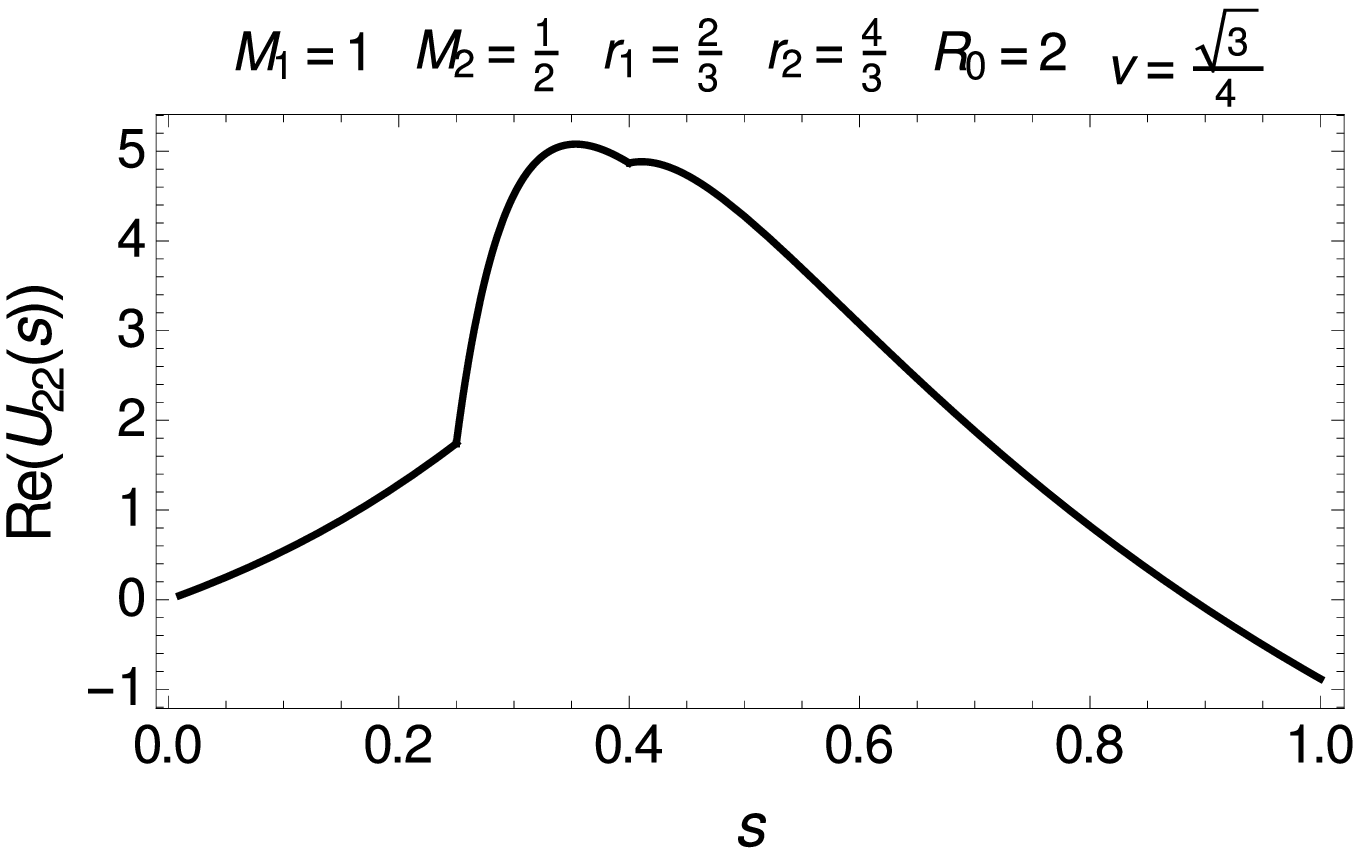}    & \includegraphics[height=4.8cm]{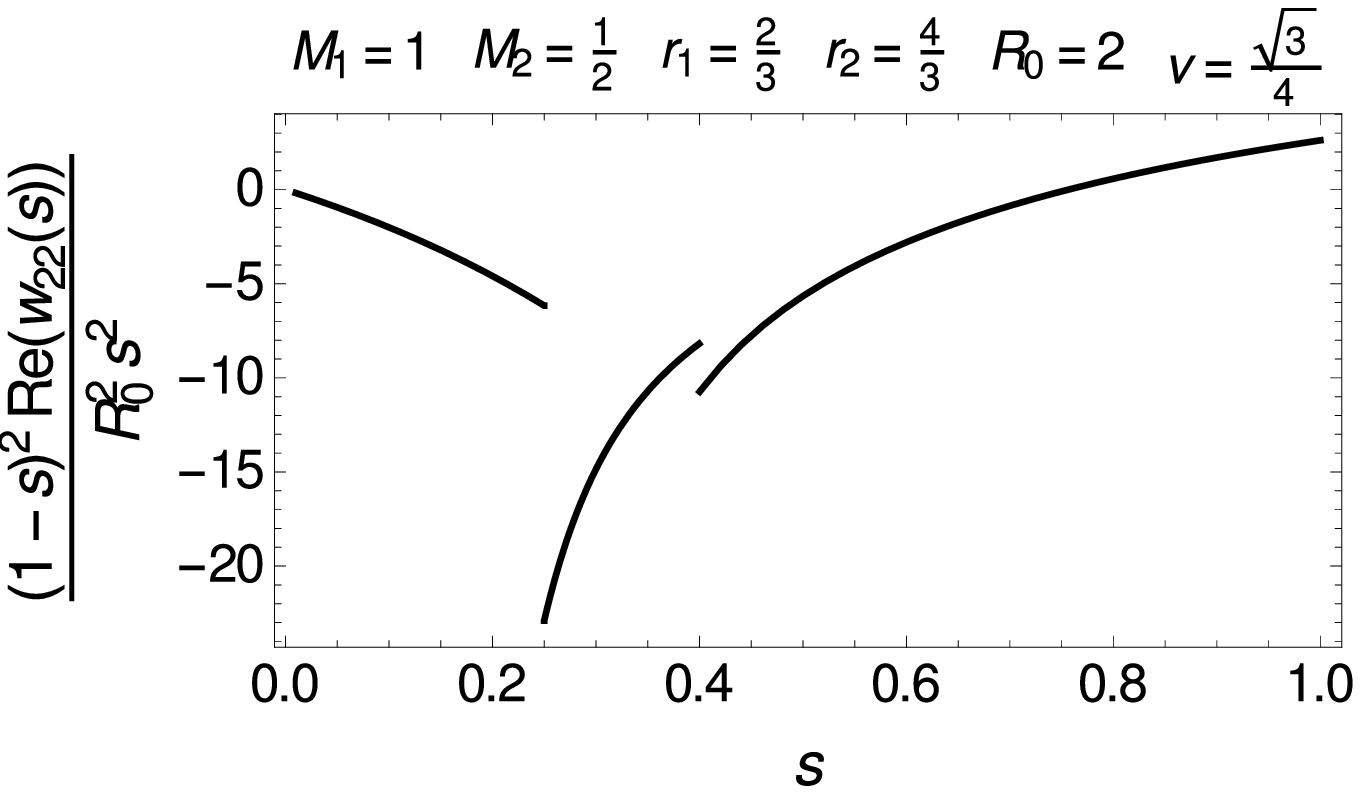}\\
(c) & (d)
\end{tabular}
\caption{Real part of some components of the metric functions ($l=m=2$) versus the compactified coordinate $s$ (see the text) for a binary system with $M_1=1/2$, $M_2=1$. The frequency of rotation is computed by means of Kepler's third law. Here $r_1$ and $r_2$ are referred to the center of mass of the system.}
\label{metric_functions}
\end{figure}
 
\begin{figure}[h!]
\begin{tabular}{c@{\hspace{0.5cm}}c}
\includegraphics[height=6.0cm]{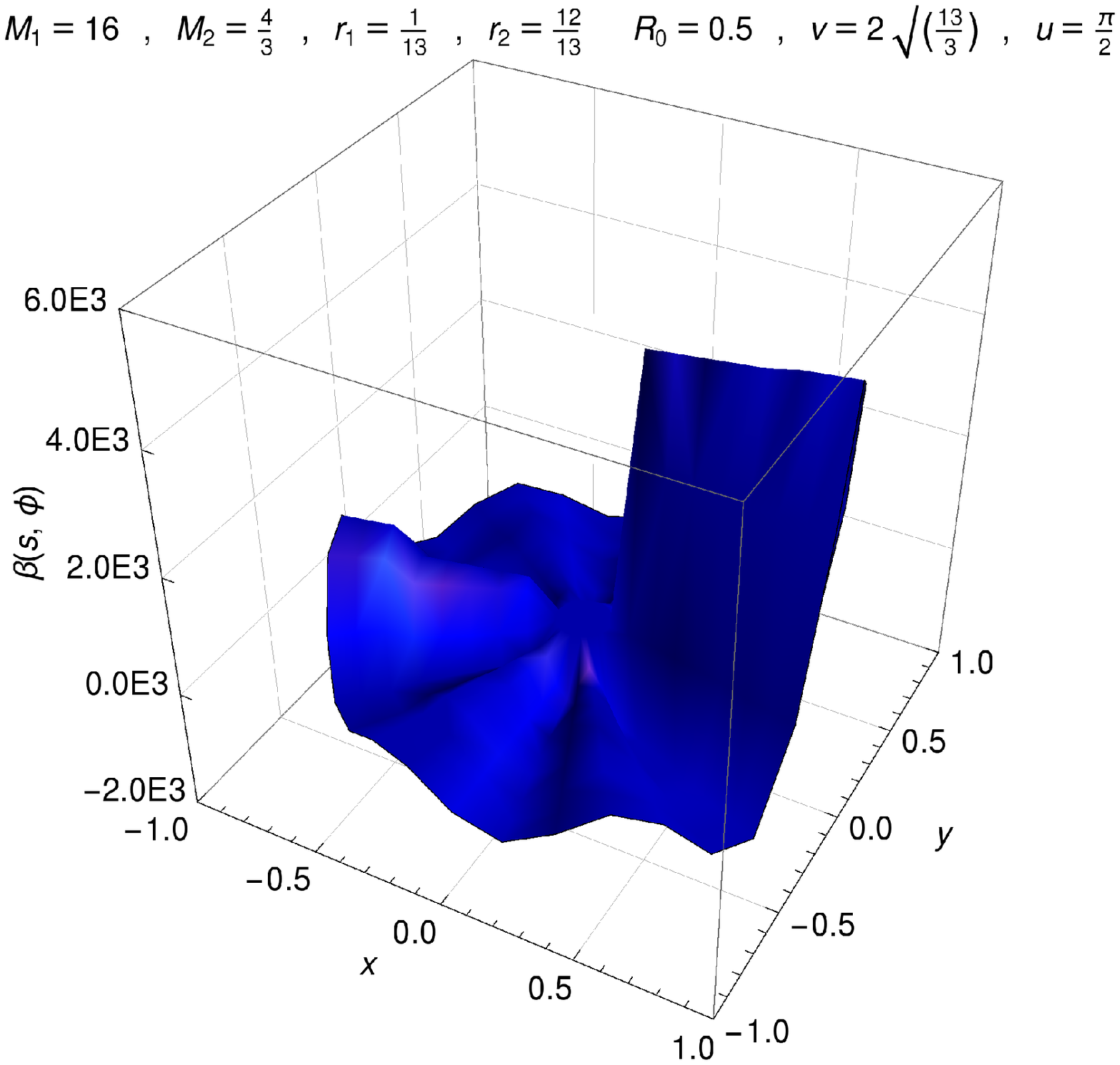} & \includegraphics[height=6.0cm]{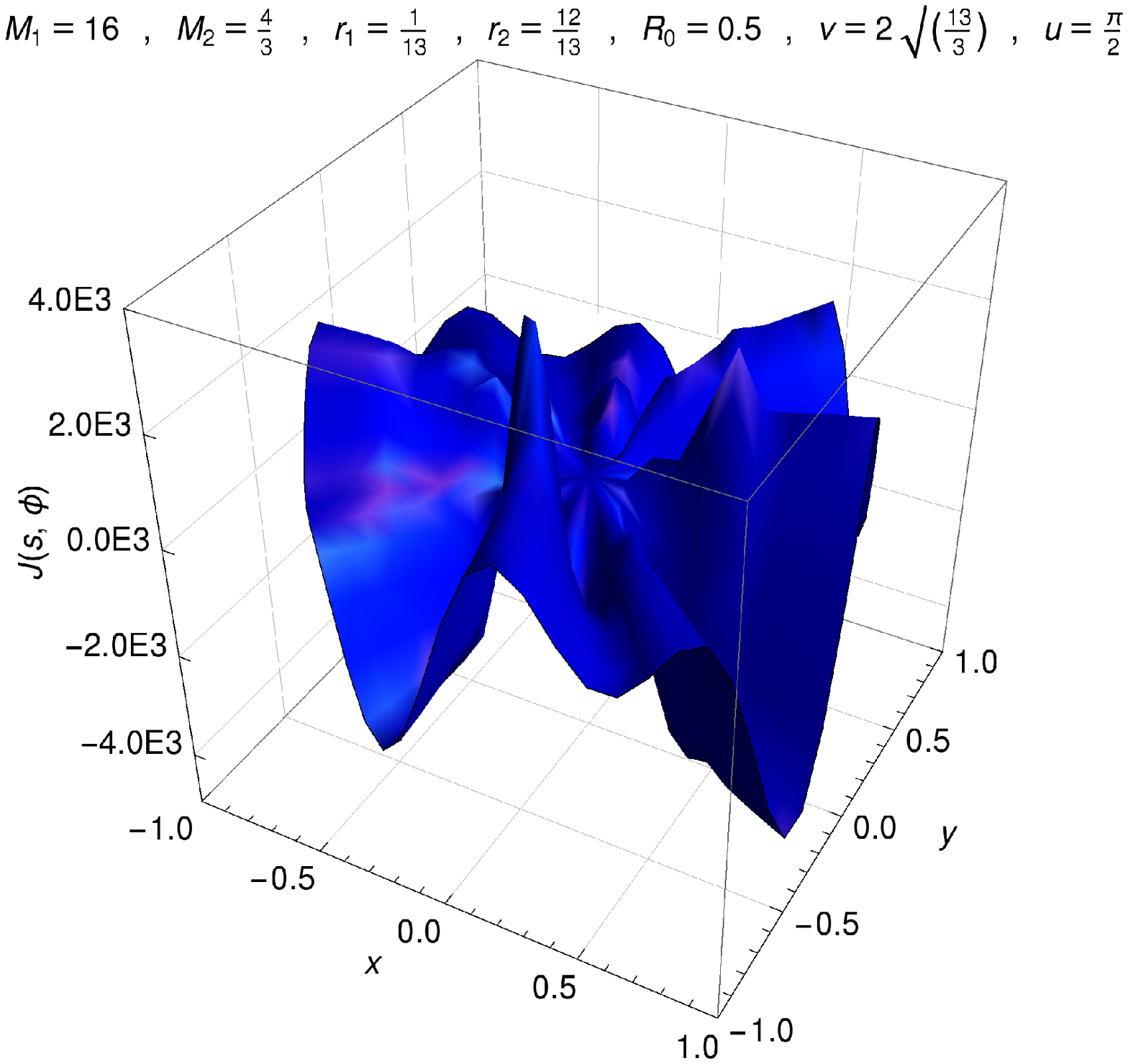}\\
(a) & (b) \\
\includegraphics[height=6.0cm]{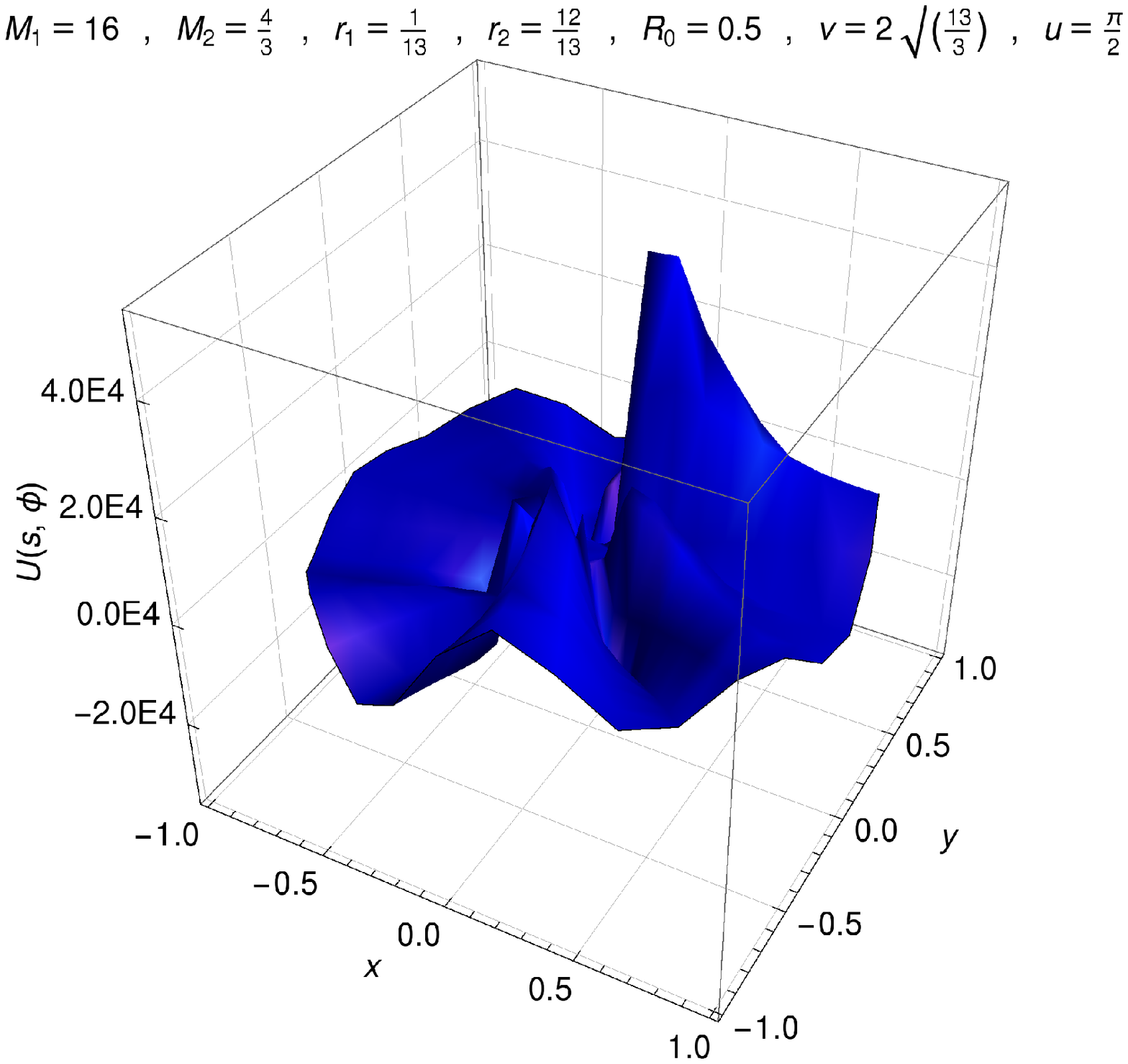}    & \includegraphics[height=6.0cm]{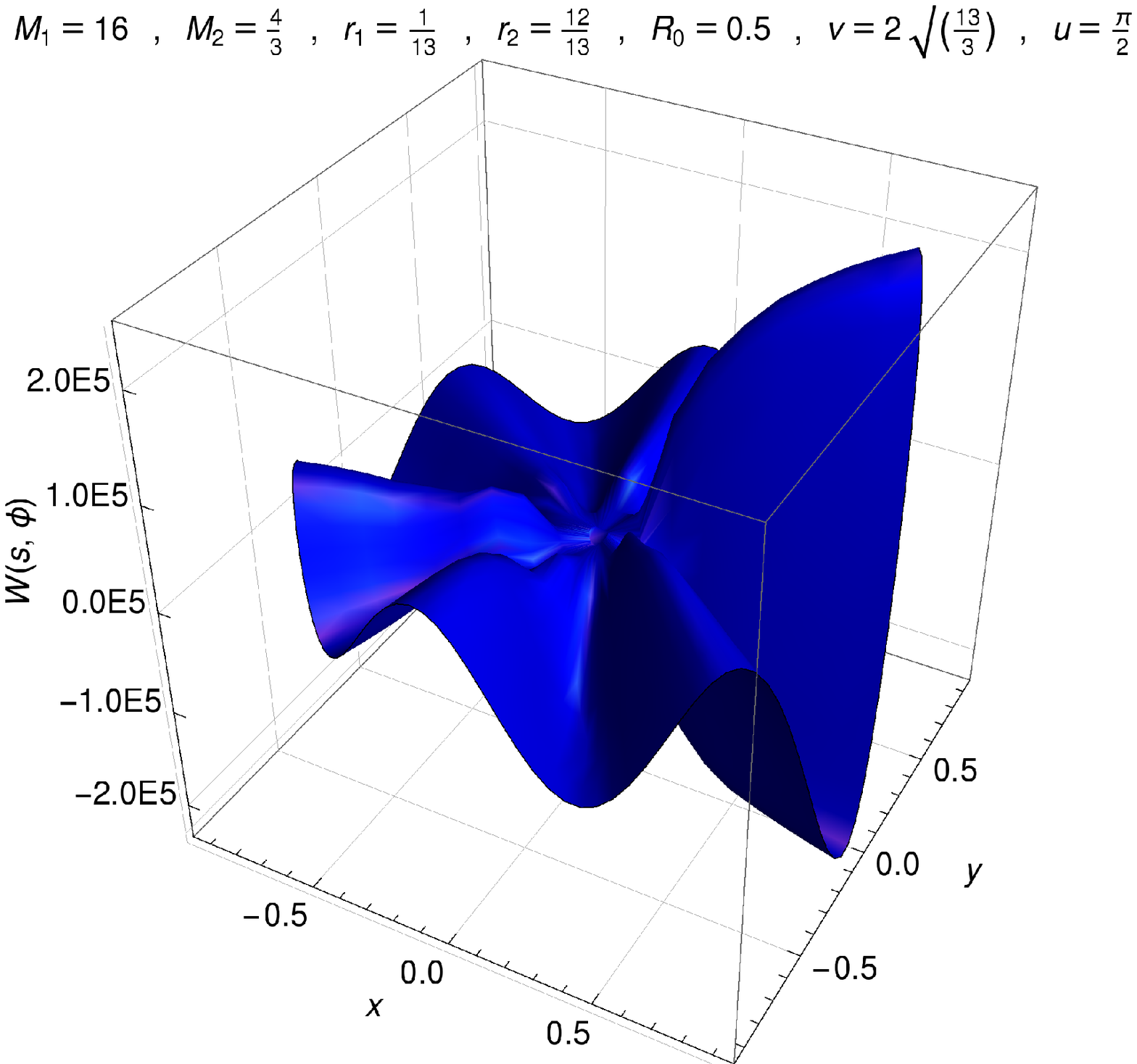}\\
(c) & (d) 
\end{tabular}
\caption{Snapshots of the metric variables as seen from the equatorial plane ($\theta=\pi/2$), as a function of $s$ and $\phi$ for $u=\pi/2$. Here $M_1=16$, $M_2=4/3$, $r_1=1/13$, $r_2=12/13$, $R_0=1/2$ and $\nu=2\sqrt{13/3}$. (a) $\beta(s,\phi)$, (b) $J(s,\phi)$, (c) $U(s,\phi)$ and (d) $W(s,\phi)=w(s,\phi)(1-s^2)/(s^2 R_0^2)$.}
\label{metric_functions1}
\end{figure}

\newpage
\noindent Here $M_1=1/2$, $M_2=1$, $R_0=2$, and the radius of each orbit is referred to the center of mass of the system, namely
\begin{equation}
r_j=\frac{\mu}{M_j}d_0\;,  \hspace{0.2cm}j=1,2\;,
\label{rdef}
\end{equation}
where $\mu$ is the reduced mass of the system and $d_0$ is the distance between the masses. The frequency of 
rotation $\nu$ is computed by means of Kepler's third law, i.e., 
\begin{equation}
\nu=\sqrt{\frac{M_1+M_2}{d_0^3}}\;.
\label{freqnu}
\end{equation} 
It is worth noting that the jumps in $\beta_{lm}$ and $w_{lm}$ functions are present at exactly $r_1$ and $r_2$, whereas for $J_{lm}$ and $U_{lm}$ only their first derivatives present discontinuities, in agreement with the boundary conditions \eref{bound_cond_6} and \eref{bound_cond_7}. 
\\To illustrate the behaviour of $\beta$, $J$, $U$ and $w$ we present them in figure \ref{metric_functions1} as a function of $s$ and $\phi$ for a particular value of the retarded time $u$. These functions are constructed by using  equation \eref{mfunctions}, and the solutions for the coefficients for each $l$ and $m$. In this case, we use $l\le 8$.
As expected, the metric functions $\beta$ and $w$ and the first derivatives of $J$ and $U$ show jumps at $(r,\theta,\phi)=(r_1,\pi/2,\nu u)$ and $(r,\theta,\phi)=(r_2,\pi/2,\nu u-\pi)$, which are just the positions of the masses, in agreement with the boundary conditions initially imposed.  
\\Note that since the first field equation for the vacuum $\beta_{,r}=0$ implies that $\beta_{lm}$ are constants along $r$, as sketched in  \fref{metric_functions}, and that $\beta$ is a gauge term for the gravitational potential, then,  $\Phi$ can be redefined as $\Phi=w/(2r)$. These facts make the choice of the frequency $\nu$ as obeying Kepler's third law completely consistent and natural.
\newpage
\section{Gravitational Radiation from the source}

Now, we proceed with the calculation of the power emitted by the binary system via gravitational wave emission. We show that the approach presented here is consistent with the well-known result 	obtained by Peters and Mathews \cite{PM63}.
\\ Following Bishop \cite{B05}, the Bondi News function in the weak field approximation is given by,
\begin{equation*}
\mathcal{N}=\lim\limits_{r\rightarrow\infty}\left(-\frac{r^2J_{,ur}}{2}+\frac{\eth^2 \omega}{2} +\eth^2\beta\right).
\end{equation*}
Substituting the metric variables given in \eref{mfunctions}, one obtains the News function for $l\ge 2$ and $-l\le m\le l$, namely
\begin{equation}
\hspace*{-1.5cm}\mathcal{N}=\lim\limits_{r\rightarrow\infty}\sum_{l,m}\Re\left(\left(-\frac{i|m|\nu r^2\left(J_{lm}\right)_{,r}}{2}+\beta_{lm}+\frac{l(l+1)J_{lm}}{4}\right)e^{i|m|\nu u}\right)\ \eth^2\ _0Z_{lm}\;.
\end{equation}
Now, substituting the coefficients of the metric variables for the exterior region, one obtains
\begin{eqnarray}
\mathcal{N}=&& -\frac{i \nu ^3 \ _2S_{21} }{\sqrt{6}}-4 i \sqrt{\frac{2}{3}} \nu^3 \ _2S_{22}  -\frac{i \nu ^4 \ _2S_{31} }{\sqrt{30} (\nu -3 i)}-\frac{8 i \sqrt{\frac{2}{15}} \nu ^4 \ _2S_{32} }{2 \nu -3 i}  \nonumber
\\&& -\frac{9 i \sqrt{\frac{3}{10}} \nu ^4 \ _2S_{33} }{\nu -i} - \frac{i \nu ^5 \ _2S_{41} }{3 \sqrt{10} \left(\nu^2-7 i \nu -14\right)}-\frac{8 i \sqrt{\frac{2}{5}} \nu ^5 \ _2S_{42} }{3 \left(2 \nu ^2-7i \nu -7\right)}   \nonumber
\\&&-\frac{81 i \nu ^5 \ _2S_{43} }{\sqrt{10} \left(9 \nu ^2-21 i \nu -14\right)} -\frac{256 i \sqrt{\frac{2}{5}} \nu ^5 \ _2S_{44} }{3 \left(8 \nu ^2-14 i \nu -7\right)}    \nonumber
\\&&+\frac{\nu ^6 \ _2S_{51} }{\sqrt{210} \left(i \nu ^3+12 \nu ^2-54 i \nu-90\right)} +\frac{16 \sqrt{\frac{2}{105}} \nu ^6 \ _2S_{52} }{4 i \nu ^3+24 \nu ^2-54 i\nu -45}  \nonumber\\
&&+\frac{27 \sqrt{\frac{3}{70}} \nu ^6 \ _2S_{53} }{3 i \nu ^3+12 \nu ^2-18 i \nu-10}+\frac{1024 \sqrt{\frac{2}{105}} \nu ^6 \ _2S_{54} }{32 i \nu ^3+96 \nu ^2-108 i \nu-45}\nonumber\\
&&+\frac{625 \sqrt{\frac{5}{42}} \nu ^6 \ _2S_{55} }{25 i \nu ^3+60 \nu ^2-54 i \nu -18} + \cdots\;, 
\label{News_all}
\end{eqnarray}
where we define the spin 2 quantity $_2S_{lm}$ as
\begin{equation}
_2S_{lm}= \frac{\left(\Re(D_{1Jlm+}e^{i|m|\nu u}) \ \eth^2\ _0Z_{lm}+ \Re(D_{1Jl-m+}e^{i|m|\nu u}) \ \eth^2\ _0Z_{l\ -m}\right)}{\sqrt{(l-1)l(l+1)(l+2)}}\;.
\end{equation}
Since the binary system is confined to a plane, then a natural choice to simplify the problem of expressing the News function, without loss of generality, is to put the masses motion in the equatorial plane $\theta=\pi/2$. This means symmetry of reflection for the density of matter and, consequently, for the spacetime. Thus, this choice restricts the components of the density, obtained from  \eref{denscomps}, to have the following form
\begin{equation}
\rho_{lm}=
\cases{
\tilde{\rho}_{lm}\frac{M_2  r_1 ^2 \delta \left(r-r_2 \right)+M_1  r_2^2 \delta \left(r-r_1
	\right)}{r_1^2 r_2^2} &  $l,m$ \ even \\ %[0.5cm]
\tilde{\rho}_{lm} \frac{M_1  r_2^2 \delta \left(r-r_1 \right)-M_2  r_1^2 \delta \left(r-r_2
	\right)}{r_1^2 r_2^2} & $l,m$ \ odd },
\label{rholm}
\end{equation}
where $\tilde{\rho}_{lm}$ are numerical constants. Therefore, for binaries %composed of components 
of different masses, 
the News function \eref{News_all} is simplified to
\begin{eqnarray}
\mathcal{N}=&& -4 i \sqrt{\frac{2}{3}} \nu^3 \ _2S_{22} -\frac{i \nu ^4 \ _2S_{31}}{\sqrt{30} (\nu -3 i)} -\frac{9 i \sqrt{\frac{3}{10}} \nu ^4 \ _2S_{33}}{\nu -i} -\frac{8 i \sqrt{\frac{2}{5}} \nu ^5 \ _2S_{42}}{3 \left(2 \nu ^2-7 i \nu -7\right)} \nonumber
\\&&  -\frac{256 i \sqrt{\frac{2}{5}} \nu ^5 \ _2S_{44}}{3 \left(8 \nu ^2-14 i \nu -7\right)} +\frac{\nu ^6 \ _2S_{51} }{\sqrt{210} \left(i \nu ^3+12 \nu ^2-54 i \nu
	-90\right)}  \nonumber
\\&& +\frac{27 \sqrt{\frac{3}{70}} \nu ^6 \ _2S_{53}}{3 i \nu ^3+12 \nu ^2-18 i \nu
	-10} +\frac{625 \sqrt{\frac{5}{42}} \nu ^6 \ _2S_{55} }{25 i \nu ^3+60 \nu ^2-54 i \nu -18} + \cdots \;.
\label{News_restricted}
\end{eqnarray}
\\When the explicit solutions are used, the News functions for the binary system take the form
\begin{eqnarray}
\mathcal{N}&=& 8\sqrt{\frac{2 \pi }{5}} \ _2L_{22} \left(\mathcal{M}_{21}+\mathcal{M}_{22}\right) \nu ^3 +\frac{1}{3} i \sqrt{\frac{\pi }{35}} \ _2L_{31} \left(\mathcal{M}_{31}-\mathcal{M}_{32}\right) \nu ^4 \nonumber\\
&-& 9 i \sqrt{\frac{3 \pi }{7}} \ _2L_{33} \left(\mathcal{M}_{31}-\mathcal{M}_{32}\right) \nu ^4 +\frac{8}{63}\sqrt{2 \pi } \ _2L_{42} \left(\mathcal{M}_{41}+\mathcal{M}_{42}\right) \nu ^5 \nonumber\\
&-& \frac{128}{9} \sqrt{\frac{2 \pi }{7}} \ _2L_{44} \left(\mathcal{M}_{41}+\mathcal{M}_{42}\right) \nu ^5
\frac{1}{180} i \sqrt{\frac{\pi }{154}} \ _2L_{51} \left(\mathcal{M}_{51} - \mathcal{M}_{52}\right) \nu ^6 \nonumber\\
&-&\frac{27}{40} i \sqrt{\frac{3 \pi }{11}} \ _2L_{53}  \left(\mathcal{M}_{51}-\mathcal{M}_{52}\right) \nu ^6
+\frac{625}{24} i \sqrt{\frac{5 \pi }{33}} \ _2L_{55} \left(\mathcal{M}_{51}-\mathcal{M}_{52}\right) \nu ^6\nonumber\\
&+&\cdots\;,
\label{News_binary}
\end{eqnarray}
where
\begin{equation}
\mathcal{M}_{lj}=M_j r_j^l(v_j^2+1)\;,
\end{equation}
and $_2L_{lm}$ are defined as
\begin{equation}
_2L_{lm}=\left(\ _2Z_{l\ -m}\Re(e^{ i |m|\nu  u}) -\Re(i e^{ i |m|\nu  u}) \ _2Z_{lm}\right)\;.
\end{equation}
Note that, as consequence of \eref{rholm}, for $M_1=M_2=M_0$ the terms with $l$ odd disappear from the News function \eref{News_binary}. Thus, as expected, one obtains immediately
\begin{eqnarray}
\mathcal{N}&=&  16 \sqrt{\frac{2 \pi }{5}} \nu ^3 M_0  r_0^2 \left(V_0^2+1\right) \ _2L_{22} +\frac{16}{63} \sqrt{2 \pi } \nu ^5 M_0  r_0 ^4 \left(V_0^2+1\right) \ _2L_{42} \nonumber \\
&&-\frac{256}{9} \sqrt{\frac{2 \pi }{7}} \nu ^5 M_0  r_0 ^4 \left(V_0^2+1\right) \ _2L_{44} +\frac{32 \sqrt{\frac{2 \pi }{13}} }{1485} \nu^7 M_0  r_0 ^6 \left(V_0^2+1\right) \ _2L_{62} \nonumber\\
&&-\frac{8192}{495} \sqrt{\frac{\pi }{195}} \nu ^7 M_0  r_0^6 \left(V_0^2+1\right) \ _2L_{64}
+\frac{2592}{5} \sqrt{\frac{2 \pi }{715}} \nu ^7 M_0  r_0 ^6\left(V_0^2+1\right) \ _2L_{66}\nonumber\\
&&+\cdots\;.
\end{eqnarray}
where $V_0$ is the physical velocity of the identical masses, which is obviously tangent to the circular orbit.
\\The energy lost by the system $dE/du$ is related to the News function, via
\begin{equation}
\frac{dE}{du}=\frac{1}{4\pi}\int_{\Omega} d\Omega \ \mathcal{N}\overline{\mathcal{N}}\;,
\end{equation} 
which results  for $M_1\ne M_2$ in
\begin{eqnarray}
\frac{dE}{du}&=&\frac{32}{5} \nu ^6 \left(\mathcal{M}_{21}+\mathcal{M}_{22}\right)^2 +\frac{2734}{315} \nu ^8 \left(\mathcal{M}_{31}-\mathcal{M}_{32}\right)^2 \nonumber \\
&+&\frac{57376}{3969} \nu ^{10} \left(\mathcal{M}_{41}+\mathcal{M}_{42}\right)^2 +\frac{4010276}{155925}\nu ^{12} \left(\mathcal{M}_{51}-\mathcal{M}_{52}\right)^2+\cdots\;.
\end{eqnarray}
Notice that the first term on the right side of the above equation is identical to
	the power lost obtained by Peters and Mathews \cite{PM63} for circular orbits, whereas the other terms stand for the octupole, hexadecapole, etc.
\section{Summary and Conclusions}
In this study we generalize a previous work by Bishop {\it et. al.} \cite{BPR11}, concerning equal mass binary systems in a Minkowski background. Here we consider the case in which the components of the binary systems have different masses, although still in circular orbits. 
\\We show that, instead of two regions, as in the case of binaries with equal components, the spacetime needs now to be separated in three regions: interior, between and outside the two world tubes. As a result, the match conditions satisfied by the coefficients in the spin-weighted spherical harmonics expansion, across the two hypersurfaces generated by the circular orbits of these two (different) masses were generalized. Also, the procedure developed here allows one to perform calculations for arbitrary values of the $l$ and $m$ modes. 
\\ It is worth stressing that one of the interesting aspects of the present study has do with the development of a procedure that can be applied in problems in which multiple layers are present. 
\\We also calculate the energy lost by the emission of gravitational waves by means of the Bondi News function. Again, we do that for arbitrary multipoles, in other words, for different values of the $l$ and $m$ modes. The interesting point here is that for different masses the emission of gravitational radiation occurs for all values (multipoles) of $l\ge 2$;
for the particular case of binary systems with equal components, the multipole terms for odd values of $l$ vanish. 
\\Last, but not least, the present study can be used as a method to extract gravitational radiation of binary systems during their periodic phase, when one considers that the system can be surrounded by a boundary, $\Gamma$, very far from the system, where a linear approximation can be applied.

\section*{Acknowledgments}
We thank the Brazilian agencies CAPES, FAPESP (2013/11990-1) and CNPq (308983/2013-0) by the financial support. We would also like to thank the referees (Dr. Casey Handmer and an anonymous) for their helpful comments and suggestions.

\appendix
\section{}
\label{appendix}
\begin{figure}[h!]
	\begin{center}
		\includegraphics[height=6cm]{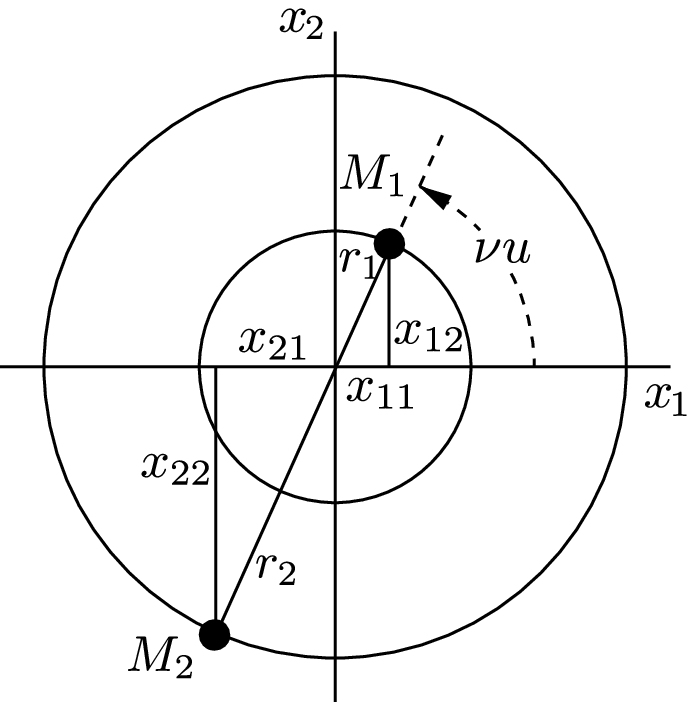}
	\end{center}
	\caption{Binary system as viewed from the top. The coordinates $x_{ai}$  of the particles are indicated, as well as the angle $\nu u$ with respect to the $x$ axis.}
	\label{fig4}
\end{figure}

It is worth recalling the main steps given by Peters and Mathews \cite{PM63} to obtain the well-known and widely used equation for the power radiated by two point masses in a Keplerian orbit.
\\In the weak field limit of the Einstein field equations, i.e., when the metric can be written as a perturbation $h_{\mu\nu}$ of the Minkowski metric $\eta_{\mu\nu}$, namely 
\begin{equation}
g_{\mu\nu}=\eta_{\mu\nu}+h_{\mu\nu}\;, \hspace{0.5cm}|h_{\mu\nu}|\ll |\eta_{\mu\nu}|\;,
\end{equation}
the power emitted by any discrete mass distribution in the limit of low velocities, as shown in \cite{PM63}, is given by
\begin{equation}
P=\frac{1}{5}\left(\tdot{Q}_{ij}\tdot{Q}_{ij}-\frac{1}{3}\tdot{Q}_{ii}\tdot{Q}_{jj}\right)\;,
\label{PM_pot}
\end{equation}
where the dots indicate derivative with respect to the retarded time $u$ and,
\begin{equation}
Q_{ij}=\sum_a m_a x_{ai}x_{aj}\;,
\label{PM_Q}
\end{equation}
where $a$ labels each particle of the system and $x_{ai}$ is the projection of the position vector of each mass along the $x$ and $y$ axes. Particularly, for a point particle binary system of different masses in circular orbits, when the Lorentz factor is considered as $\gamma=1$,  one can write
\begin{equation}
x_{a1}=r_a \cos(\nu u-\pi \delta_{a2})\;, \hspace{0.5cm}x_{a2}=r_a \sin(\nu u-\pi \delta_{a2})\;,\hspace{0.5cm} a=1,2\;,
\label{PM_x}
\end{equation}
as shown in \fref{fig4}
\\Here, $r_a$ is given by \eref{rdef}, $\nu$ by \eref{freqnu}, and the Kronecker delta distinguishes each  particle. Then, the components of $Q_{ij}$ reads
\begin{eqnarray}
Q_{ij}=\pmatrix{\mu d_0^2 \cos^2(\nu u) & \mu d_0^2 \sin(\nu u)\cos(\nu u) \cr 
	\mu d_0^2 \sin(\nu u)\cos(\nu u) & \mu d_0^2 \sin^2(\nu u)}\;,
\label{PM_Q1}
\end{eqnarray}
thus,
\begin{eqnarray}
\tdot Q_{ij}=\pmatrix{ 4 \nu^3 \mu d_0^2\sin (2 \nu  u) & -4 \nu^3 \mu d_0^2\cos (2 \nu  u)\cr 
	-4 \nu^3 \mu d_0^2\cos (2 \nu  u) & -4 \nu^3 \mu d_0^2 \sin (2 \nu  u)}\;.
\label{PM_Q1}
\end{eqnarray}
Finally, substituting the above equation in \eref{PM_pot}, one obtains
\begin{equation}
P=\frac{32}{5} \mu ^2 \nu ^6 d_0^4 = \frac{32 {m_1}^2 {m_2}^2 (m_1+m_2)}{5 d_0^5}\;,
\end{equation}
where Kepler's third law is used in the last equality.
\section*{References}
\bibliographystyle{iopart-num.bst}
%\bibliography{References}

\begin{thebibliography}{10}
\expandafter\ifx\csname url\endcsname\relax
  \def\url#1{{\tt #1}}\fi
\expandafter\ifx\csname urlprefix\endcsname\relax\def\urlprefix{URL }\fi
\providecommand{\eprint}[2][]{\url{#2}}

\bibitem{B14}
Blanchet L 2014 {\em Living Reviews in Relativity\/} {\bf 17}
  \urlprefix\url{http://www.livingreviews.org/lrr-2014-2}

\bibitem{BBM62}
Bondi H, van~der Burg M~G~J and Metzner A~W~K 1962 {\em Proc. Phys. Soc. A\/}
  {\bf 269} 21--52

\bibitem{S62}
Sachs R~K 1962 {\em Proc. Phys. Soc. A\/} {\bf 270} 103--126

\bibitem{BGLW96}
Bishop N~T, G\'omez R, Lehner L and Winicour J 1996 {\em Phys. Rev. D\/} {\bf
  54}(10) 6153--6165

\bibitem{BGLMW97}
Bishop N~T, G\'omez R, Lehner L, Maharaj M and Winicour J 1997 {\em Phys. Rev.
  D\/} {\bf 56}(10) 6298--6309

\bibitem{BGLMW99}
Bishop N~T, G\'omez R, Lehner L, Maharaj M and Winicour J 1999 {\em Phys. Rev.
  D\/} {\bf 60}(2) 024005

\bibitem{G01}
G\'omez R 2001 {\em Phys. Rev. D\/} {\bf 64}(2) 024007

\bibitem{BGLMW05}
Bishop N~T, G\'omez R, Lehner L, Maharaj M and Winicour J 2005 {\em Phys. Rev.
  D\/} {\bf 72}(2) 024002

\bibitem{GBF07}
G\'omez R, Barreto W and Frittelli S 2007 {\em Phys. Rev. D\/} {\bf 76}(12)
  124029

\bibitem{RBLTS07}
Reisswig C, Bishop N~T, Lai C~W, Thornburg J and Szilagyi B 2007 {\em Class.
  Quantum Grav.\/} {\bf 24} S327

\bibitem{OR11}
de~Oliveira H~P and Rodrigues E~L 2011 {\em Class. Quantum Grav.\/} {\bf 28}
  235011

\bibitem{C13}
Cao Z 2013 {\em Int. J. Mod. Phys. D\/} {\bf 22} 1350042

\bibitem{RBP13}
Reisswig C, Bishop N and Pollney D 2013 {\em Gen. Relativ. Gravit.\/} {\bf 45}
  1069--1094

\bibitem{HS15}
Handmer C~J and Szil\'agyi B 2015 {\em Class. Quantum Grav.\/} {\bf 32} 025008

\bibitem{B05}
Bishop N~T 2005 {\em Class. Quantum Grav.\/} {\bf 22} 2393

\bibitem{RBPS10}
Reisswig C, Bishop N~T, Pollney D and Szil\'agyi B 2010 {\em Class. Quantum
  Grav.\/} {\bf 27} 075014

\bibitem{BPR11}
Bishop N, Pollney D and Reisswig C 2011 {\em Class. Quantum Grav.\/} {\bf 28}
  155019

\bibitem{K12}
{Kubeka} A~S 2012 {\em Journal of Modern Physics\/} {\bf 3} 1503--1515

\bibitem{PM63}
Peters P~C and Mathews J 1963 {\em Phys. Rev.\/} {\bf 131}(1) 435--440

\bibitem{TBRSCKS13}
Taylor N~W, Boyle M, Reisswig C, Scheel M~A, Chu T, Kidder L~E and Szil\'agyi B
  2013 {\em Phys. Rev. D\/} {\bf 88}(12) 124010

\bibitem{GPLPW97}
G\'omez R, Lehner L, Papadopoulos P and Winicour J 1997 {\em Class. Quantum
  Grav.\/} {\bf 14} 977

\bibitem{NP66}
{Newman} E~T and {Penrose} R 1966 {\em J. Math. Phys.\/} {\bf 7} 863--870

\bibitem{GMNRS66}
{Goldberg} J~N, {Macfarlane} A~J, {Newman} E~T, {Rohrlich} F and {Sudarshan}
  E~C~G 1967 {\em J. Math. Phys.\/} {\bf 8} 2155--2161

\bibitem{W83}
{Winicour} J 1983 {\em Journal of Mathematical Physics\/} {\bf 24} 1193--1198

\bibitem{ZGHLW03}
Zlochower Y, G\'omez R, Husa S, Lehner L and Winicour J 2003 {\em Phys. Rev.
  D\/} {\bf 68}(8) 084014

\end{thebibliography}
\providecommand{\noopsort}[1]{}\providecommand{\singleletter}[1]{#1}%
\providecommand{\newblock}{}

\end{document}